\documentclass[twocolumn,english,aps,prb,floatfix,preprintnumbers,showpacs,amsfonts,amssymb,superscriptaddress]{revtex4-1}

\usepackage[T1]{fontenc}
\usepackage[utf8]{inputenc}
\usepackage[english]{babel}

\usepackage{amsmath}
\usepackage{amssymb}
\usepackage{graphicx}

\begin{document}

\title{Magnetocaloric effect and~frustrations in~one-dimensional magnets}

\author{A. V. Zarubin}
\email{Alexander.Zarubin@imp.uran.ru}
\affiliation{Institute of Metal Physics, Ural Division, RAS, Russia, Ekaterinburg, S.~Kovalevskaya str. 18}

\author{F. A. Kassan-Ogly}
\affiliation{Institute of Metal Physics, Ural Division, RAS, Russia, Ekaterinburg, S.~Kovalevskaya str. 18}

\author{M. V. Medvedev}
\affiliation{Institute of Electrophysics, Ural Division, RAS, Russia, Ekaterinburg, Amundsen str. 106}

\author{A. I. Proshkin}
\affiliation{Institute of Metal Physics, Ural Division, RAS, Russia, Ekaterinburg, S.~Kovalevskaya str. 18}

\begin{abstract}
In this paper, we investigated the magnetocaloric effect (MCE) in one-dimensional magnets with different types of ordering in the Ising model, Heisenberg, $XY$-model, the standard, planar, and modified Potts models. Exact analytical solutions to MCE as functions of exchange parameters, temperature, values and directions of an external magnetic field are obtained. The temperature and magnetic field dependences of MCE in the presence of frustrations in the system in a magnetic field are numerically computed in detail.
\end{abstract}


\maketitle

\section*{Introduction}

Magnetocaloric effect is a phenomenon associated with a change in the temperature of a magnet with adiabatic excitation ($dS=0$) of an external magnetic field. When the magnetization of the system changes, the value of the MCE can be calculated from the relationship
\begin{equation}
\left(\frac{\partial T}{\partial h}\right)_S=-\frac TC \left(\frac{\partial S}{\partial h}\right)_T= -\frac TC \left(\frac{\partial M}{\partial T}\right)_h,
\label{eq:MCE0}
\end{equation}
where $T$ is the temperature, $S$ is entropy, $C$ is specific heat, $M$ is magnetization, $h$ is an external magnetic field. The second definition appears to be the most convenient because it is determined in terms of the quantity that can be measured in experiment.

Importantly, the study of MCE is a topical problem of condensed matter physics. The study of magnetocaloric effect in various magnets performed in conjunction with studies of their other magnetic and thermal properties provide information about the relationship of these features.

Since MCE at $T\to 0$ can reach very high values, it can be used for achieving ultra-low temperatures upon adiabatic demagnetization of the objects.

From the technological applications point of view, a large MCE is the most important property for currently intensive searching the materials suitable for usage as cooling agents in magnetic refrigerators.

The theoretical description and prediction of materials with high values or even giant MCE creates real opportunities for the effective selection of the construction for working magnetic refrigeration devices.

Of particular interest is the study of MCE in the systems, magnetic structure of which can be interpreted as one-dimensional one. A number of such spin systems exhibit unusual magnetic properties due to the peculiarities of spatial or magnetic structure, and, particularly, the presence of  frustrations~\cite{Sadoc:1999,Diep:2004,Zhitomirsky:2004,Kassan-Ogly:2010}. The study of such systems is convenient within the one-dimensional spin models, as prototype due to possibility of obtaining the exact analytical solutions.

The usage of 1D-models appears to be justified for a number of compounds (e.g., cerium monochalcogenides and monopnictides), since these prototype models give correct qualitative description of real three-dimensional magnets, when appropriate rescalings of material parameters are taken into account.

\section{Model and basic formulas}

In this case, developing the theory of MCE for one-dimensional magnetic systems is based on the model Hamiltonian that takes into account the interaction between nearest and next-nearest neighbors, as well as the external magnetic field, and that is defined as
\begin{equation}
\mathcal{H}=-J\sum_i (\sigma _i,\sigma_{i+1})-J'\sum_i (\sigma _i,\sigma_{i+2})-\sum_i (h,\sigma_i),
\label{eq:H}
\end{equation}
where $J$ is the parameter of exchange interaction between the magnetic moments of nearest neighbors (exchange integral) located at the sites $i$ and $i+1$; $J'$ is the  parameter of exchange interaction between the magnetic moments of the second neighbors, located at the sites $i$ and $i+2$; $h$ is the value of an external magnetic field; $\sigma_i$ is the  $z$-projection of the spin operator stochastically taking the values~$\pm 1$.

Equation (2) allows us to write the Hamiltonian in a general form for Ising, Heisenberg, $XY$-model, standard and planar Potts models, as well as modified 6-, 8- and 12-states Potts models~\cite{Kassan-Ogly:2011,Kassan-Ogly:2013}, bounding the consideration only by the systems with localized magnetic moments.

This approach makes it possible to solve the problem of MCE description in a large variety of models, specifying only the configurations of the magnetic system and the parameters of the exchange interaction and the external magnetic field.

Obtaining the exact solutions to such models is most convenient by the Kramers--Wannier transfer-matrix method~\cite{Kassan-Ogly:2010,Kassan-Ogly:2013}. In this method, the partition function is determined by the maximum eigenvalue of the transfer-matrix, and the whole set of all thermodynamic characteristics of the system (free energy, internal energy, entropy, specific heat, magnetization, susceptibility, and also the magnetocaloric effect) is solely expressed in terms of the maximum eigenvalue.

In this paper were obtained exact analytical expressions for the maximum eigenvalue $\lambda $ of the transfer-matrix as a function of temperature, the exchange parameters $J$, $J'$, and the value and direction of the magnetic field $h$, for various models, and the resulting expression for MCE (\ref{eq:MCE0}) in all the models in the form of
\begin{equation}
MCE=
\frac{
\lambda \frac{\partial \lambda}{\partial h} + T\lambda \frac{\partial }{\partial T}\frac{\partial \lambda}{\partial h} - T\frac{\partial \lambda}{\partial h}\frac{\partial \lambda}{\partial T}
}{
2\lambda \frac{\partial \lambda}{\partial T} - T\left(\frac{\partial \lambda}{\partial T}\right)^2 + T\lambda\frac{\partial ^2\lambda}{\partial T^2}
}.
\label{eq:MCE}
\end{equation}

\section{MCE in para-, ferro- and antiferromagnets}

\begin{figure}[htb]
\begin{center}
\includegraphics[width=\columnwidth]{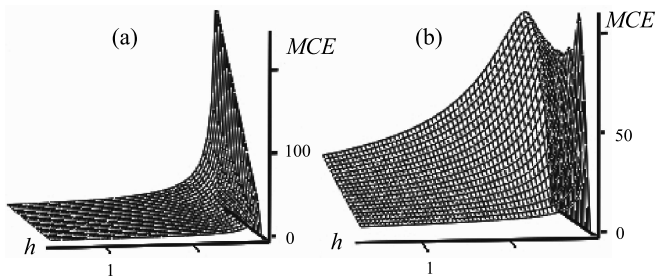}
\end{center}
\caption{Typical dependences of MCE on temperature and field in a paramagnet (a) and a ferromagnet, the interaction between nearest neighbors (b)}
\label{fig:1}
\end{figure}

We considered one-dimensional magnetic systems with different types of magnetic ordering. In these magnets, the MCE is significantly different (Fig.~\ref{fig:1}).

Our results show that in any paramagnetic spin system ($J=J'=0$) with localized magnetic moments, irrespective of the model, the lattice type, the space dimension, and also at arbitrary spin value the MCE, as a function of temperature and external magnetic field, is always positive and is equal to $T/h$ (Fig.~\ref{fig:1}a).

In the one-dimensional spin system with ferromagnetic interaction between the nearest neighbors MCE (as a function of temperature and external magnetic field) is positive and reaches  maximum values when both the temperature and magnetic field  are tending to zero (Fig.~\ref{fig:1}b).

\begin{figure}[htb]
\begin{center}
\includegraphics[width=\columnwidth]{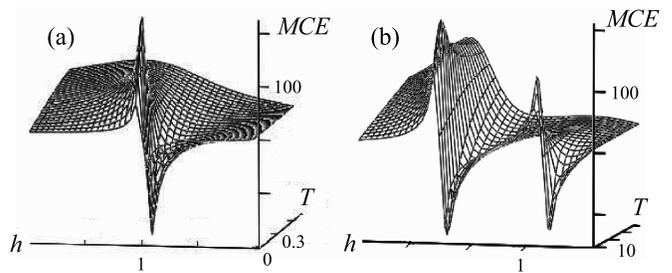}
\end{center}
\caption{Temperature and field dependence of MCE: (a)~-- nearest-neighbor interaction, (b)~-- competing antiferromagnetic interactions, several frustration fields in the Ising model}
\label{fig:2}
\end{figure}

Calculations have shown that in all the models with ferromagnetic interaction the qualitative behavior of MCE, as a function of temperature and external magnetic field is similar, and only the scale of the effect is different.

In the lattice with antiferromagnetic exchange interaction, the behavior of MCE is substantially determined by the presence of frustrations in the system. Such a magnet has at least one frustration field near which MCE reaches giant values (at low temperatures). Such behavior of MCE in antiferromagnets differ significantly from that in the cases of paramagnets and magnets with ferromagnetic exchange between the lattice spins (Fig.~\ref{fig:2}a).

\begin{figure}[htb]
\begin{center}
\includegraphics[width=\columnwidth]{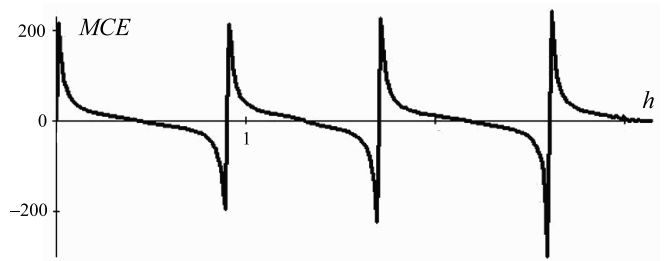}
\end{center}
\caption{Field dependence of MCE at low temperatures in a magnet with four frustration fields. 4-state Potts model with competing antiferromagnetic interactions}
\label{fig:3}
\end{figure}

In the Ising model with arbitrary spin value $s$, with taking into account the nearest-neighbor interaction, the frustration field itself, entropy and magnetization at the frustration field at $T=0$ are equal to
\begin{gather}
h_{\text{F}}=-2sJ,\quad S=\frac{1+\sqrt{1+8s}}{2},\notag\\ M=\left(\frac{2s-1}{4}\right)+\left(\frac{2s+1}{4}\right)\frac{1}{\sqrt{1+8s}}.
\label{eq:F}
\end{gather}
In the presence of competing interactions with either $J<0$ and $J'<0$ or $J>0$ and $J'<0$ in both these models, depending on the direction of the magnetic field several frustration fields may appear, near each of which the change in sign of MCE and its giant growth occur (Fig.~\ref{fig:2}b). The frustrations generate anomalous behavior of MCE with sign changes and giant growth nearby the frustration fields (Fig.~\ref{fig:3}).

\section*{Summary}

Thus, the conditions for the existence of the giant MCE in one-dimensional magnetic systems are obtained. Moreover, the magnetic properties of such systems with local magnetic moments are substantially determined by the presence of frustrations.

MCE can reach maximum values at low temperatures in the magnets with frustrations, and the value is the greater the closer the external magnetic field to the frustration field is.

From our considerations it follows that the most effective method for achieving low temperatures is the cyclic switching of an external magnetic field about each of the frustration field. The results provide some predictive possibilities in selection of magnetic-ordering materials as a cooling agent for magnetic refrigeration machines.

\bigskip

This work was supported by the Presidium of RAS, Project no.~12-P-2-1041 and Project of Integration and Basic Research of Ural Division of RAS, no.~12-I-2-2020.

\bibliography{mce2014}

\begin{thebibliography}{6}%
\makeatletter
\providecommand \@ifxundefined [1]{%
 \@ifx{#1\undefined}
}%
\providecommand \@ifnum [1]{%
 \ifnum #1\expandafter \@firstoftwo
 \else \expandafter \@secondoftwo
 \fi
}%
\providecommand \@ifx [1]{%
 \ifx #1\expandafter \@firstoftwo
 \else \expandafter \@secondoftwo
 \fi
}%
\providecommand \natexlab [1]{#1}%
\providecommand \enquote  [1]{``#1''}%
\providecommand \bibnamefont  [1]{#1}%
\providecommand \bibfnamefont [1]{#1}%
\providecommand \citenamefont [1]{#1}%
\providecommand \href@noop [0]{\@secondoftwo}%
\providecommand \href [0]{\begingroup \@sanitize@url \@href}%
\providecommand \@href[1]{\@@startlink{#1}\@@href}%
\providecommand \@@href[1]{\endgroup#1\@@endlink}%
\providecommand \@sanitize@url [0]{\catcode `\\12\catcode `\$12\catcode
  `\&12\catcode `\#12\catcode `\^12\catcode `\_12\catcode `\%12\relax}%
\providecommand \@@startlink[1]{}%
\providecommand \@@endlink[0]{}%
\providecommand \url  [0]{\begingroup\@sanitize@url \@url }%
\providecommand \@url [1]{\endgroup\@href {#1}{\urlprefix }}%
\providecommand \urlprefix  [0]{URL }%
\providecommand \Eprint [0]{\href }%
\providecommand \doibase [0]{http://dx.doi.org/}%
\providecommand \selectlanguage [0]{\@gobble}%
\providecommand \bibinfo  [0]{\@secondoftwo}%
\providecommand \bibfield  [0]{\@secondoftwo}%
\providecommand \translation [1]{[#1]}%
\providecommand \BibitemOpen [0]{}%
\providecommand \bibitemStop [0]{}%
\providecommand \bibitemNoStop [0]{.\EOS\space}%
\providecommand \EOS [0]{\spacefactor3000\relax}%
\providecommand \BibitemShut  [1]{\csname bibitem#1\endcsname}%
\let\auto@bib@innerbib\@empty
\bibitem [{\citenamefont {Sadoc}\ and\ \citenamefont
  {Mosseri}(1999)}]{Sadoc:1999}%
  \BibitemOpen
  \bibfield  {author} {\bibinfo {author} {\bibfnamefont {J.-F.}\ \bibnamefont
  {Sadoc}}\ and\ \bibinfo {author} {\bibfnamefont {R.}~\bibnamefont
  {Mosseri}},\ }\href {\doibase 10.1017/CBO9780511599934} {{\selectlanguage
  {english}\emph {\bibinfo {title} {Geometrical frustration}}}}\ (\bibinfo
  {publisher} {Cambridge University Press},\ \bibinfo {address} {New York},\
  \bibinfo {year} {1999})\BibitemShut {NoStop}%
\bibitem [{\citenamefont {Diep}(2004)}]{Diep:2004}%
  \BibitemOpen
  \bibinfo {editor} {\bibfnamefont {H.~T.}\ \bibnamefont {Diep}},\ ed.,\
  \href@noop {} {{\selectlanguage {english}\emph {\bibinfo {title} {Frustrated
  spin systems}}}}\ (\bibinfo  {publisher} {World Scientific},\ \bibinfo
  {address} {New Jersey},\ \bibinfo {year} {2004})\BibitemShut {NoStop}%
\bibitem [{\citenamefont {Zhitomirsky}\ and\ \citenamefont
  {Honecker}(2004)}]{Zhitomirsky:2004}%
  \BibitemOpen
  \bibfield  {author} {\bibinfo {author} {\bibfnamefont {M.~E.}\ \bibnamefont
  {Zhitomirsky}}\ and\ \bibinfo {author} {\bibfnamefont {A.}~\bibnamefont
  {Honecker}},\ }\href {\doibase doi:10.1088/1742-5468/2004/07/P07012}
  {\bibfield  {journal} {\bibinfo  {journal} {J. Stat. Mech. Theor. Exp.}\
  }\textbf {\bibinfo {volume} {2004}},\ \bibinfo {pages} {P07012} (\bibinfo
  {year} {2004})}\BibitemShut {NoStop}%
\bibitem [{\citenamefont {Kassan-Ogly}\ and\ \citenamefont
  {Filippov}(2010)}]{Kassan-Ogly:2010}%
  \BibitemOpen
  \bibfield  {author} {\bibinfo {author} {\bibfnamefont {F.~A.}\ \bibnamefont
  {Kassan-Ogly}}\ and\ \bibinfo {author} {\bibfnamefont {B.~N.}\ \bibnamefont
  {Filippov}},\ }\href {\doibase 10.3103/S1062873810100394} {\bibfield
  {journal} {\bibinfo  {journal} {Bulletin of the RAS: Physics}\ }\textbf
  {\bibinfo {volume} {74}},\ \bibinfo {pages} {1452} (\bibinfo {year}
  {2010})}\BibitemShut {NoStop}%
\bibitem [{\citenamefont {Kassan-Ogly}\ and\ \citenamefont
  {Filippov}(2011)}]{Kassan-Ogly:2011}%
  \BibitemOpen
  \bibfield  {author} {\bibinfo {author} {\bibfnamefont {F.~A.}\ \bibnamefont
  {Kassan-Ogly}}\ and\ \bibinfo {author} {\bibfnamefont {B.~N.}\ \bibnamefont
  {Filippov}},\ }\href {\doibase 10.4028/www.scientific.net/SSP.168-169.427}
  {\bibfield  {journal} {\bibinfo  {journal} {Solid State Phenom.}\ }\textbf
  {\bibinfo {volume} {168-169}},\ \bibinfo {pages} {427} (\bibinfo {year}
  {2011})}\BibitemShut {NoStop}%
\bibitem [{\citenamefont {Kassan-Ogly}\ \emph {et~al.}(2013)\citenamefont
  {Kassan-Ogly}, \citenamefont {Medvedev}, \citenamefont {Proshkin},\ and\
  \citenamefont {Zarubin}}]{Kassan-Ogly:2013}%
  \BibitemOpen
  \bibfield  {author} {\bibinfo {author} {\bibfnamefont {F.~A.}\ \bibnamefont
  {Kassan-Ogly}}, \bibinfo {author} {\bibfnamefont {M.~V.}\ \bibnamefont
  {Medvedev}}, \bibinfo {author} {\bibfnamefont {A.~I.}\ \bibnamefont
  {Proshkin}}, \ and\ \bibinfo {author} {\bibfnamefont {A.~V.}\ \bibnamefont
  {Zarubin}},\ }\href {\doibase 10.3103/S1062873813100122} {\bibfield
  {journal} {\bibinfo  {journal} {Bulletin of the RAS: Physics}\ }\textbf
  {\bibinfo {volume} {77}},\ \bibinfo {pages} {1245} (\bibinfo {year}
  {2013})}\BibitemShut {NoStop}%
\end{thebibliography}%
\bibliographystyle{apsrev4-1}

\end{document}